\documentclass[pra,aps,groupedaddress,twocolumn,floatfix,nofootinbib]{revtex4-2}

\usepackage{hyperref}
\usepackage{multirow}
\usepackage{mathrsfs,amsmath} 
\usepackage{verbatim}
\usepackage{lipsum}
\usepackage{dsfont}
\usepackage{footnote}
\usepackage{amsfonts}
\usepackage{mathtools}
\usepackage{amsthm}
 \usepackage{graphicx}
 \usepackage[table,xcdraw]{xcolor}
	
\usepackage[euler]{textgreek}
\usepackage[normalem]{ulem}

\definecolor{shadecolor}{rgb}{0.8,0.9,1}
\DeclareDocumentCommand{\Tr}{m m O{\big}}{{\rm Tr}_{\:\!{#1}}#3({#2}#3)}

\newcommand{\vir}{``}

\begin{document}
\title{Creative and geometric times in physics, mathematics, logic, and philosophy}

\author{Flavio Del Santo}
\affiliation{Group of Applied Physics, University of Geneva, 1211 Geneva, Switzerland}
\affiliation{Constructor University, Geneva, Switzerland}

\author{Nicolas Gisin}
\affiliation{Group of Applied Physics, University of Geneva, 1211 Geneva, Switzerland}
\affiliation{Constructor University, Geneva, Switzerland}

\date{\today}

\begin{abstract}
\noindent 
We  distinguish two different concepts of time that play a role in physics: \textit{geometric  time} and \textit{creative time}. The former is the time of deterministic physics and merely parametrizes a given evolution. The latter is instead characterized by real change, i.e. novel information that gets created when a non-necessary event becomes determinate in a fundamentally indeterministic physics. This allows one to give a naturalistic characterization of the present as the moment that separates the potential future from the determinate past. We discuss how these two concepts find natural applications in classical and intuitionistic mathematics, respectively, and in classical and intuitionistic logic, as well as how they relate to the well-known A- and B-theories in the philosophy of time. We acknowledge that we do not offer here a unified concept or a new philosophy of time. However, we contend that none of the existing philosophical accounts fully integrate both the geometric and creative concepts of time.

\end{abstract}

\maketitle

\section{Introduction}

There is almost nothing that we perceive so ubiquitously as the occurrence of time. And yet our most successful physical theories  still struggle to make sense of this concept in an unequivocal way. Actually, modern physics has relegated  time to play a less and less special role \cite{smolin2013time}. However, in the words of I. Prigogine,  $\vir$no formulation of the laws of nature that does not take into account this constructive role of time can ever be satisfactory" \cite{prigogine1997end}. 

To address this fundamental problem, we distinguish here two different concepts of time, which we call \textit{geometric time} and \textit{creative time}, respectively. We show that those stem from our fundamental assumption of physics as being either deterministic or indeterministic at the fundamental, ontological level and that they both seem to contribute to our understanding of natural phenomena.
In particular, we will see that geometric time is the parametric time appearing in the (deterministic) equations of motions of physics, just a coordinate that, together with the three spatial directions, labels fixed events in a geometrical block-universe.

On the other hand, creative time arises from the assumption that certain events, which are fundamentally indeterminate, become determinate. Our general approach is admittedly influenced by quantum mechanics, which as a matter of fact brought the notions of fundamental indeterminacy and indeterminism into physics. More abstractly, we consider the possibility of \textit{ontic indeterminacy}: a metaphysical feature of a physical quantity that exhibits “worldly imprecision” \cite{miller2021worldly, miller2024classical, arroyo2025incomplete, torza2023indeterminacy}. An ontologically indeterminate reading of the quantum state is, arguably, a widely held option (setting aside nonlocal hidden‐variable theories or many‐worlds). Although not a logical necessity, Heisenberg uncertainty together with violations of Bell inequalities provide strong reasons to regard quantum indeterminacy as more than a reasonable hypothesis. Moreover, we have shown in detail that, even at the classical level, if one adopts a finiteness‐of‐information principle, there is a consistent way to interpret classical theory as fundamentally indeterminate \cite{gisin2020real, del2019physics, del2021indeterminism, del2023prop}. Throughout the paper, we therefore take seriously the working hypothesis of ontic (sometimes called fundamental) indeterminacy.

In general, indeterminacy is a kinematic property of the state describing such a quantity. There is nothing subjectively uncertain here; in our naturalistic view, Nature simply has not yet fully determinate all objects.
Indeterminism, on the other hand, refers to the possibility in Nature of having different potential (mutually exclusive) future states, given the present state and the laws of physics.
Note that ontic indeterminacy in Nature is sufficient (though not necessary) for the existence of multiple potential future states, i.e., for indeterminism.

Indeterminacy can be formalized in terms of the information existing about, for example, a certain physical property. In fact, we contend that the world is composed of material entities -- e.g., particles and fields -- but these do not exist \textit{simpliciter}. What exists is structure: information about the arrangement of things, their relations, their positions, etc. This information is objective, \textit{existing} in nature independently of any observer. It is physically instantiated in the degrees of freedom of material systems: a physicalist interpretation in the spirit of Landauer’s dictum that information is physical \cite{landauer}.

More information can be created when a potential future is actualized, thereby reducing indeterminacy. It is the actualization of potentialities -- when new information that previously did not exist is created -- that makes creative time tick. This is in line with the ideas of mathematician A.~Connes, who states: "It's quantum effervescence that generates the passage of time, not the other way around." \cite{connes2013theatre}.

This idea of creative time also resonates to some extent with the positions of the philosopher H.~Bergson, well known for having confronted Einstein about the nature of time, who stated: $\vir$Time is what prevents everything from being given all at once. [...] It must therefore be development. Wouldn’t it then be the vehicle of creation and choice? Would time’s existence not prove that things are indeterminate? Would time not be this very indeterminacy?" \cite{bergson2021pensee}. However, while vaguely inspired by Bergson, as physicists we twist the story around and adopt a more explicitly naturalistic view: it is the natural change in metaphysical indeterminacy (i.e., the creation of novel information and the passage from a potentiality to an actual property) that defines our creative time.
 We can thus say that, while geometric time ``passes'', creative time ``processes'' or ``occurs'', as the consequence of changes in the world.\footnote{In the literature one often finds that time $\vir$unfolds" or that time $\vir$flows" to indicate that time is associated with something that really changes, and thus with creative time. Here we prefer to use the more active terms $\vir$time occurs" or $\vir$time processes", because a film already shot does unfold, but nothing new is created there; whereas to flow is a somewhat misleading term, for it recalls the concept of speed, which already presupposes time.}

\begin{figure*}[]
\includegraphics[width=17cm]{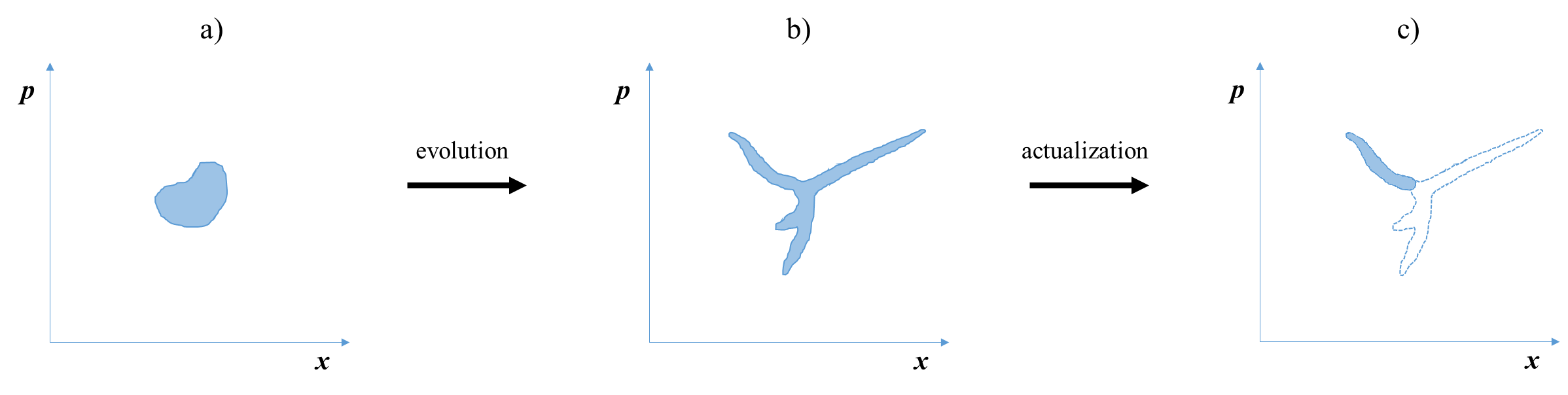}
\caption{\small{Typical evolution of a classical chaotic system in phase space. The initial state (a) has some ontic indeterminacy, i.e., the position and momentum are in this case not determinate further than the finite region depicted in figure. When evolved through the deterministic equations of motions (b), the indeterminacy spreads (while preserving the same volume), leading to multiple potential future evolutions, i.e. to indeterminism. However, only one of these potential evolutions will actualize (c), while ruling out the other previously potential futures (dashed); it is this process of actualization that makes creative time process.}}
\label{chaos}
\end{figure*}

Starting from physics (Sect. \ref{phys}), we apply these two concepts of time to different disciplines, showing that this distinction gives rise to natural parallels in mathematics (classical versus intuitionistic; Sect. \ref{math}), helps clarify the logic of scientific propositions (Sect. \ref{logic}), and it contributes further to the metaphysics of time (so-called A-theory versus B-theory; Sect. \ref{phil}). More specifically, the approach based on consecutive actualizations of potentialities, that is when creative time is at work, leads to a naturalistic description of the difference between past, present, and future.

We do not claim that the account we present here constitutes a new philosophy of time, nor that it resolves enduring debates, such as that between phenomenalism and scientism, regarding our perception or the intrinsic nature of time. Rather, we aim to emphasize the distinction between two different concepts of time, which we believe arise naturally from physical science and have parallels in various other disciplines.

\section{Two times in physics}
\label{phys}
In order to illustrate different concepts of time in physics, let us start from an example. Let us ask the following  scientific question: what will be the weather like, say on the field hockey  pitch of Geneva, exactly in one month from now? 
One can envision several procedures to reply to this question. On the one hand, it is possible to use the equations of motion of all the involved molecules of air, water vapor, etc., to calculate their evolution and so predict the weather in Geneva in a month's time (thereby applying an algorithm, which is by definition a set of finite instructions). On the other hand, one can wait exactly one month, let the process develop, and finally observe the weather in Geneva. In-between, one could spend one week to collect more data and only then start computing the evolution of all air molecules. Let us concentrate on the two first mentioned extremal procedures. The question that we ask here is: Are these two procedures talking about the same unique concept of time, or are there multiple ones that play a role in physics? We contend that there are (at least) two different concepts of time, as exemplified by the two procedures above.

Before elaborating on that, we ought to distinguish between different fundamental interpretations of physics, which become especially  relevant in dealing with complex systems like the one determining the weather in our example. The standard narrative, (i) has it that classical physics is deterministic, i.e. that each state is mapped one-to-one to all other possible states at all times (with time being the parameter that appears in the equations of motion). We would like to stress that this is arguably based on a popular misconception, while, in fact, most physical theories can be regarded as indeterministic \cite{earman1986primer}. In fact, the belief that classical physics is deterministic is based on two independent ingredients: the first is that since the equations of motion are ordinary differential equations, their solutions generally have a unique continuation in the parameter time,\footnote{In exceptional cases, classical equations of motion of even simple systems do not have a unique (deterministic) solution, such as in the example of Norton's dome \cite{norton2003causation}. Moreover, in highly complex systems --  such as Lagrangian fluid particles in high Reynolds-number turbulence -- the fact that deterministic trajectories are not unique is rather common (see \cite{eyink2020renormalization} and references therein).} given appropriate initial conditions; the second is that the initial conditions exist fully determinate with infinite precision (i.e., a state is represented by a dimensionless mathematical point in phase space, in turn described by an n-tuple of real numbers). Only taking these two assumptions together justifies the purported fundamental determinism of classical physics. Another possibility, (ii) is to keep the idea that the system is fundamentally deterministic, but realistically it is not possible to know the initial conditions with infinite precision. This becomes most relevant in chaotic systems (like the weather in our example) in which the future behavior of the considered system highly depends on the infinitesimal digits of the initial conditions. In this case one can only statistically predict what are the possible future states compatible with the partial knowledge of the initial conditions. Finally, (iii) one can instead uphold the more radical assumption that the initial conditions (alternatively, the state of a system) simply \textit{do not exist} determinate with infinite precision, but there is an\textit{ ontic indeterminacy} in nature (for a collection of  works on this see, \cite{drossel2015relation, del2019physics, gisin2020real, gisin2021indeterminism, del2021indeterminism, del2023prop}; see also \cite{calosi2019quantum} and references therein for recent developments on metaphysical indeterminacy). In this case, although the equations of motion may remain unchanged, the system does not undergo a necessary evolution, but multiple future states are possible (perhaps with a determinate objective tendency or propensity; see \cite{del2023prop}), i.e., it is the ontic indeterminacy that leads to  fundamental indeterminism (see Fig. \ref{chaos}). In what follows, when referring to indeterminacy we will have this kind of ontic property in mind and its consequent indeterminism.

Let us now go back to analyze the concepts of time that are at work in the example of the weather in Geneva in a month. The first procedure, under the ideal assumption of determinism (i), merely makes use of time as a parameter that simply indicates the coordinate of a certain evolution (e.g., sunny weather in one month from the start of the computation). Together with the three spatial coordinates (i.e., in Geneva in our example), this kind of time forms a manifold that contains all the events: a four-dimensional block universe picture where past, present, and future are all on the same metaphysical footing (this is also known as \textit{Eternalism} in the philosophy of time; see Sect. \ref{phil}). This can thus be called \textit{geometric time}, the time that most physicists have come to think of as fundamental, both from classical physics and even more from relativity theory where time becomes “just" another dimension like space.\footnote{Though with a different signature in the metric.} In this sense, time is a relational label that allows one to position an event (as earlier or later than to another point of reference), but there is no new information created in the world. 

On the other hand, let us consider the case of ontic indeterminacy in the physical state (iii). There, the actual development of a non-necessary physical process -- such as the evolution of a highly chaotic system like the one that determines the weather, or of a quantum system -- involves, in our opinion, a time that really $\vir$processes". That is, a concept of time that sets an actual separation of the future from the past (with the present at their interface). And it does so when new information that was not existing before gets created -- i.e., even having access to the whole existing information of the initial conditions (even of the whole state of the universe) \textit{before}, would not determine all the future physical properties of a system \textit{afterwards}. Again, we stress that we are not talking here about epistemic uncertainty, but rather we explicitly take a physicalist stance on information: it quantifies something objectively encoded in the degrees of freedom of the universe. In this sense, information is \textit{created} when non-necessary physical events obtain, independently of any observer. Note that -- as in quantum physics -- even given the pure state of a system -- i.e., in Butterfield's words, “a system’s maximal (or ‘complete’) set of intrinsic (or ‘possessed’) properties” \cite{butterfield2005determinism} -- this does not fully determine the future evolution (see also Ref.~\cite{miller2021worldly}). Here, \textit{complete} means that a state contains the maximal information presently available \vir in nature" about each potential property, i.e.\ the collection of all propensities. Yet only one potentiality, among its alternatives, eventually actualizes. The concept of time that arises from this can thus be called \textit{creative time}.\footnote{The distinction between geometric and creative time was sketched in Refs. \cite{Gisin2017, del2021relativity, del2019physics, santo2023open} and previously also in Ref. \cite{horwitz1973relativistic} where the creative time was called $\vir$historical time".} On a similar note, G. Ellis remarked that $\vir$The time that is the present at this instant will be the past at the next instant. [...] It is this fundamental feature of time that is not encapsulated by today's theoretical physics" \cite{ellis2008flow} (see also \cite{dolev2018physics}).

Let us emphasize that the concept of creative time has nothing to do with language, nor with subjective or psychological processes; it ticks due to objective processes that happen by themselves in nature. Accordingly, there are two types of events: necessary events (e.g., when two geometric worldlines cross each other) and non-necessary events which occur when potentialities become actual. This is exemplified, for instance, by the case of a single photon impinging on an array of detectors—eventually, only one detector clicks.

So, coming back to the example of the weather in Geneva in a month, one can carry out either of the envisioned procedures to determine the outcome -- (i) compute the state of the weather in a month using the relevant laws of physics, or (ii) wait and directly observe the weather. Both these procedures take some time. In practice, to carry out the first procedure, one maps the time interval that it takes to perform the computation $\Delta t'$ to the time appearing in the equations of motion. That is, to the interval $\Delta t=t_f-t_i$, where $t_i$ is the parametric time associated to the initial state and $t_f$ is the parametric time of the considered final state (i.e., the moment when we want to know the weather in Geneva). Note that both $\Delta t'$ and $\Delta t$ are geometric times. For this to be a prediction, the (deterministic) computation should last  $\Delta t'< \Delta t=1$ month, and return as an outcome the weather in Geneva at $t_f$, say $\vir$sunny".\footnote{In principle the time for the “prediction" could take longer than the time interval to observe the outcome. In such a case, one would anyways ask for consistency of the calculation afterwards.}

One can also think of more subtle questions about the weather, such as $\vir$will there be (anytime in history) 700 consecutive days of rain in Geneva?". If the  complex system that forms the weather is fully deterministic, answering this question seems only a matter of geometric time, although the answer may remain undecided for any finite value of geometric time. One can run the computation of the dynamics of the system and at some point find that the answer is $\vir$yes", if one finds the 700 consecutive days of rain, otherwise the algorithm will keep searching without ever halting (however, this question might be more complex, see Sect. \ref{math}).

Note that  $\Delta t'$ can be sped up by a more efficient computation, namely, this depends on how fast one $\vir$runs" through the parameter time as it appears in the equations of motion (this depends on the used algorithm and, physically, one can really think of the speed of the head of a Turing machine in reading the tape). Moreover, one can use any other effects that allow one to manipulate geometric time, such as time dilation in relativity that modifies the interval of geometric time  $\Delta t'$ (see Subsect. \ref{rela}).  So, the map $\Delta t\rightarrow \Delta t'$ is a one-to-many function. 

In contrast, in the process involving creative time, then one has no choice but to wait a month and see what happens, because new information is created along the way through a series of indeterministic processes. So, the initial and the final states are not deterministically mapped to each other. Creative time is required when an actualization event of the values of physical variables that were previously indeterminate occurs. It is the process of change from indeterminate to determinate  -- with an associated creation of novel information -- that brings about this “occurrence" of time: creative time ticks when new information is created.

\section{Two times in Mathematics}
\label{math}

Although it is not common to discuss time in mathematics at all, the problem of the two different kinds of time introduced above has almost a perfect counterpart in mathematics.     
Let us now consider a standard mathematical object, e.g. a real number. Typical real numbers -- in fact, almost all of them in a mathematical sense -- are uncomputable, i.e. for each of these numbers there exists no algorithm -- which is by definition a \textit{finite} list of instructions -- that outputs all their digits (they have infinite Kolmogorov complexity) \cite{gisin2021indeterminism}. On the contrary, the (irrational) numbers that we usually consider as prototypical examples of reals, such as $\sqrt{2}$ or $\pi$, are all computable (the ratio between the diagonal of a square to its edge is a simple algorithm that outputs all the digits of $\sqrt{2}$ and similarly the ratio of the circumference to the diameter of a circle for $\pi$).
Obviously, all the rational numbers are also computable: they can be directly written down or compressed (e.g., in the case of repeating decimals) in a finite string. 

The standard way of regarding mathematics, the so-called classical-mathematics, is a form of Platonism which posits that, among the other mathematical entities, real numbers exist with their infinite series of digits, although there is in general no way to even label and thus grasp them (since there is no algorithm that generates them  \cite{gisin2020mathematical, gilles}). Hence, classical Platonistic mathematics is a timeless language, which per se is not a bug, but it becomes problematic when mathematics is elevated to the language of science, i.e.,  is used to describe the physical world.

However, there are alternative approaches known as constructive mathematics, of which one of the most prominent is intuitionism (see \cite{handbookconstr, posy2020mathematical, bentzen2025intuitionism} and references therein). Therein, mathematical entities are not given at once, but rather are processes that develop  in time (digits get \textit{created} continuously, one after the other).\footnote{\label{newInfo}More precisely, it is not necessarily new digits that come into existence, but the new information reduces the indeterminacy of intuitionistic real numbers    \cite{gilles}.}  This explicitly provides mathematics with a concept of passage of time (and it would be only at infinite time -- at the end of time, so to say \cite{posy2020mathematical} -- that the mathematical entities, such as real numbers, are completed into the ones defined by classical mathematics). Indeed, one should stress that \vir the dependence of intuitionism on time is essential: statements can become provable in the course of time and therefore might become intuitionistically valid while not having been so before". \cite{sep-intuitionism}.

The initiator of intuitionism, L.E.J. Brouwer,  envisioned intuitionistic mathematics as \vir having its origin in the perception of a move of time" \cite{brouwer1981brouwer}. However, for him mathematics was the activity of the mind of a $\vir$creative subject", i.e., an idealized mind or mathematician, who is responsible for this progressive process of creation in time of, for instance, the digits of an uncomputable real numbers \cite{brouwer1981brouwer}. Note that, for Brouwer, time is merely internal, i.e., it is not physical time at all -- neither geometric nor creative in the sense discussed here (see \cite{bentzen2024brouwer}). Several authors, however, have distanced themselves from this controversial concept of a creative \textit{subject}. In particular, one of us (N.G.) has put forward the idea that the digits of a typical real number are generated by a \textit{\vir natural" random number generator} (see also footnote \ref{newInfo}), i.e., a natural process -- as opposed to the action of an agent -- that creates new information in the world by changing a fundamentally indeterminate bit into a determinate one  \cite{gisin2021intuitionism}.\footnote{For more works that relate constructive mathematics to physics, see \cite{gisin2020mathematical} and \cite{fletcher2002constructivist}.} This version of intuitionism, which departs from the standard view (see e.g., \cite{posy2020mathematical, van2006brouwer}), could thus be labeled $\vir$objective" or $\vir$naturalistic" intuitionism, as characterized in \cite{bentzen2025naturalistic}.  In this way, typical real numbers become a graspable concept directly linked to creative time, that is, to the change from the indeterminate to the determinate happening in natural phenomena. Hence, naturalistic intuitionism, is a tensed mathematical language. 

To think about computable real numbers, one just needs geometric time. In fact, it is possible to think that the full information about those numbers is contained in the (finite) algorithm that defines them, i.e., in their initial conditions. For instance, one can ask what is the 43800th decimal place of $\pi$ and the answer is given by running  an algorithm that outputs the digits of $\pi$ and picking its 43800th decimal. The answer happens to be a 9, and it is easily computable because it is not necessary to go through all the previous digits of $\pi$ to compute 43800th one. There are so-called \textit{digit-extraction algorithms} that allow one to directly compute the nth digit of $\pi$ \cite{plouffe2022formula}.\footnote{The most well-known of such algorithms is perhaps the Bailey-Borwein-Plouffe formula \cite{bailey1997rapid}, that allows to directly compute the hexadecimal digits of $\pi$. Digit-extraction algorithms, also in base 10, are known for several other irrational computable real numbers, such as for $e$.} This can be seen as a further evidence that computable (though irrational) numbers are already fully determinate and do not require creative time.
The point is that since this is found through a deterministic outcome, the answer to this kind of question is fully contained in the algorithm, which can be run more or less fast, giving the ability to manipulate this $\vir$mathematical geometric time". This exactly resembles the weather example in Sect. \ref{phys} for the deterministic evolution of the physical system that forms the weather. 

In contrast, if one considers instead a typical, i.e., uncomputable, real number, things are different. There are no algorithms compressing the information of that number. The only existing $\vir$algorithm" is the number itself (more precisely there is no finite list of operations that generates the number, but the only way to indicate the number is to list its infinite digits). Each next digit is generated by a genuinely random natural process. Therefore, asking what the 43800th digit of such a number requires waiting through all 43800 instances of the creation of the $\vir$next digit" (see footnote \ref{newInfo}); that is, it requires $\vir$mathematical creative time".

In intuitionism, these numbers are generated in time by what are usually called non-lawlike choice sequences, i.e., sequences of computable numbers  with new bits (or digits) created as time processes.
\cite{van2006brouwer, posy2020mathematical, bentzen2025intuitionism,  moschovakis2020intuitionistic, niekus2025individual}.\footnote{In what follows, we will consider that uncomputable numbers are generated by non-lawlike sequences, in contrast to computable numbers that are generated by lawlike sequences.} In our naturalistic version of intuitionism \cite{bentzen2025naturalistic}, fresh bits get generated by a natural random number generator, i.e., a process occurring in nature that is able to produce true randomness, and a choice sequence $\alpha(n)$ is the outcome of a computable function that takes as an argument $\alpha(n-1)$ --  where $n$ is the time index -- and all the random bits previously generated (assumed to be independent and identically distributed).\footnote{Note that although the computable function deterministically produces outcomes given suitable inputs, the inputs themselves are random—thus rendering the output indeterminate.} Typical, uncomputable real numbers can be regarded as generated through this natural process of creation of new information as creative time processes.

This has a connection to the mathematical model of indeterminate physics that was introduced in Ref. \cite{del2019physics}. There, the standard real number are substituted by so-called \textit{finite information quantities} (FIQs), which only encode the propensity of the digits of a physical quantity to actualize. FIQs can be regarded as a particular species of choice sequences, where the next generated bit is biased by taking the majority vote of the values taken by the previous $k$ bits, with $k$ being an odd integer.
In this framework, the propensities emerging in FIQs stem from the inherent randomness of $50\%-50\%$ bit choices within these sequences. If a strong majority of recent $k$ bits is 1, it is likely that this trend will persist in the near future. This likelihood is captured as a propensity, constrained to rational values. (see \cite{gisin2021intuitionism} for a detailed discussion).\footnote{To fully recover the comparison between physics and mathematical intuitionism, one might expect a discussion extending beyond the notion of choice sequences. The present work, however, focuses on a more foundational stage: establishing the meaningfulness of formulating physical theory using choice sequences, particularly in light of a constructive reconsideration of time.}

Finally, one can also ask a question of the kind $\vir$are there 700 consecutive sevens in the digits of a number.\footnote{This example is borrowed from C. Posy \cite{posy2020mathematical}.} If that number is uncomputable, then one has no choice but to wait and see. However, even for computable numbers, like e.g. $\pi$, the question is interesting. The situation is similar to the example of the weather -- i.e., whether there will be 700 consecutive  days of rain if one assumes determinism. One can, in fact, think of programming two types of software: The first, $S_1$, is exactly the one we used in the physical example of the weather, i.e., the systematic search of 700 consecutive sevens in $\pi$; on the other hand, the second software, $S_2$ outputs all the theorems that can be derived within Peano arithmetic and it halts should it find the negation of the conjecture that there are  700 consecutive sevens in $\pi$. If the conjecture is true, $S_1$ will find such a sequence of 700 sevens and therefore halt after a finite (geometric) time. If the conjecture is provably false (within Peano's arithmetic), $S_2$  will halt. Interestingly, if neither halts, then the statement is false (otherwise, $S1$ would halt), but this is unprovable within Peano's arithmetic (otherwise, $S_2$ would halt).\footnote{Given that the lifetime of the universe is sufficient for carrying out the calculation.} Moreover, for the sake of completeness, there is also the fourth logical possibility that both $S_1$ and $S_2$ halt, a case which would prove Peano's arithmetic inconsistent. This would be an example of G\"odel's celebrated theorem.
Note that mathematicians don't use software like $S_1$ and $S_2$, but use their creativity to find shortcuts to analyse such statements, i.e., de facto they use creative time. 
In the case of the weather in Geneva, if one assumes that the system is indeterministic, like uncomputable numbers, then there is no choice but wait and see. However, if one trusts the equations describing the weather evolution and the fully determinate initial conditions, then one could run analogues of $S_1$ and $S_2$, as for computable numbers. Note however, that this assumes one can compress the initial conditions into a finite algorithm (hence into a computer which is necessarily finite).

We would like to emphasize that one of the key features of intuitionism is that, while introducing time and finiteness into mathematics, it still preserves the concept of the continuum. This makes intuitionism a suitable mathematical framework for physics, as most physical theories rely on continuity (e.g., Lorentz transformations in relativity require the continuum). However, the intuitionistic continuum is \vir sticky" or \vir viscous" \cite{posy2020mathematical} because, at any finite time, choice sequences are still ongoing, and one cannot select a specific point from the continuum, as it is not yet determinate whether the sequence will converge to above or below or on the specific point. This is the origin of Brouwer’s famous continuity theorem that states that all total functions , i.e., functions defined for every element of its domain, are continuous \cite{gisin2021intuitionism}.

Let us acknowledge that this distinction between geometric and creative time in mathematics distances our view from the most accepted position in intuitionism, such as that of C. Posy \cite{posy2020mathematical} or M. van Atten \cite{van2018creating, van2022dummett}, who consider all numbers, whether computable or not, as generated by new information that comes about, so that they all require creative time in our parlance. Exactly how these approaches to intuitionism diverge, philosophically, is a question we have only recently begun to investigate with B.~Bentzen \cite{bentzen2025naturalistic} and that will require more systematic work.

\section{Logic}
\label{logic}
We have seen that creative time emerges, both in physics and in mathematics, from indeterminacy, i.e.,  from the possibility of having new information that comes into existence. Another way to investigate the role of time and indeterminacy is to consider
propositions about physical properties and analyze their logical structure.


In this logical framework, one formulates physical properties as propositions; for example, $p(R)$ denotes the proposition ``the system has property $R$''. The aim of this approach is to assign truth values to physical propositions and to combine them using logical connectives. In particular, $p(R)$ is true, at the empirical level, if and only if an ideal (dichotomic) measurement of the corresponding property of the system would  yield the outcome $R$ with certainty.\footnote{By ``dichotomic'' we mean that the measurement can only answer, ideally, whether the system has property $R$ or not.} In a classical physical theory, this criterion can be implemented by identifying $R$ with a region of phase space and checking whether the system’s state is entirely contained in $R$.

Two central principles uphold in classical logic are the law of non-contradiction,
$\neg(p \wedge \neg p)$, and the law of the excluded middle (LEM),
$p \vee \neg p$. We now examine how different notions of time influence the logical structure of propositions about physical properties, and how these two fundamental principles are affected in relation to underlying ontologies, whether deterministic or indeterministic.

Determinism in logic can be defined as follows: ``If $p$ is true then it has always been the case that $p$ will be true'', for all propositions $p$. Indeterminism is just its negation: ``$\exists p$ such that $p$ is true and it has \textit{not} always been the case that $p$ will be true''~\cite{surowik2024tense}. 
In what follows, we discuss two main arguments that, in our opinion, hold in logic when applied to physics: (i) the law of non-contradiction requires a (geometric) time label for both determinism and indeterminism; and (ii) the law of the excluded middle fits best with deterministic logic (and hence with geometric time), whereas indeterminism (related to creative time) fits best with the rejection of the LEM.

\textbf{(i) \textit{Law of non-contradiction}}\\
Since physical states evolve in time, propositions about physical properties, such as $p(R)$ above, must reflect this temporality. In fact, if one asserts \emph{simpliciter} -- without any temporal qualification -- that ``the system has property $R$'' and also that ``the system does not have property $R$'', one is led to an apparent contradiction. For a system whose state changes in time, each statement may be correct at different instants; hence, without a time index, both $p(R)$ and $\neg p(R)$ would be true, thereby violating the principle of non-contradiction.

A natural remedy is to supply a temporal index. Once one writes ``the system has property $R$ at time $t$'' and ``the system does not have property $R$ at time $t$'', the principle of non-contradiction is respected, since at any given time only one of these propositions can be true (the pure state of a system is unique at any given time). By contrast, if the propositions concern different instants, $t_1 \neq t_2$, both may be true without inconsistency. Thus the principle of non-contradiction applies only to propositions referring to the same time, a point already emphasized by Kant~\cite{kovac2020immanuel}.

Hence, time is necessary for logical consistency when dealing with propositions referring to physical systems. In deterministic physics, this time can simply be the parameter appearing in the equations of motion, i.e., geometric time. In indeterministic physics, by contrast, one must consider the logical structure of propositions in which creative time plays a role.

\textbf{(ii) \textit{Law of the excluded middle}}\\
The law of the excluded middle (LEM) sits naturally within a logically deterministic framework, in which the truth values of all propositions are fixed and constant in time. By contrast, in a fundamentally indeterministic physics---where future contingencies are not yet determinate and become so only at a later moment of creative time---logic models in which the LEM may fail are more naturally motivated, as already noted by Aristotle (see \cite{whitaker1996aristotle}, Ch.~9). In such a setting, the truth values of propositions about the future need not be determinate in the present. To capture this situation, one may adopt non-classical logical frameworks that explicitly accommodate indeterminacy.

In the remainder of this section, we discuss some of these approaches, which illustrate the failure of the LEM and provide formal frameworks aligned with indeterminacy and the associated notion of creative time. We stress that we do not intend here either to give a systematic review of these suitable logical systems, nor to advocate for any specific one. Our main aim is to give notable examples from the literature that align with our goal of introducing temporality into logic in a way that the LEM does not hold, for the description of indeterministic physics. We will now show that this can be achieved in different ways, such as by introducing temporality explicitly into the syntax (like in Prior's tense logic where a future operator is introduced as a new logical connective), or solely into the semantics (like in Kripke models in intuitionistic logic).

A first possibility for dealing with indeterminate propositions would be to adopt a multivalued logic -- with more truth values than just true and false, for example an ``indeterminate'' value -- as in the influential work of J.~{\L}ukasiewicz \cite{lukasiewicz1968determinism}. This approach, however, raises other foundational difficulties, such as the controversial definition of negation, and will therefore not be discussed further here.

On the other hand, tense logic -- when combined with an open-future and certain semantics -- offers a way to maintain a two-valued logic while allowing a failure of the LEM for future contingents. No third truth value is introduced: propositions are always either true or false relative to a given time and a given future branch. However, in an open-future setting, propositions about the future are evaluated with respect to multiple admissible branches. As a result, a future-tense proposition may be true on some branches and false on others, and thus the LEM need not hold globally for future contingents, even though bivalence is preserved locally on each branch.\footnote{Strictly speaking, tense logic does not reject the LEM at the syntactic level; the failure arises at the semantic level under branching-time (open-future) interpretations, which we assume throughout.}

This perspective was first formalized by A.~Prior (see \cite{sep-future-contingents} and references therein) within a modal framework introducing temporal operators such as the future operator $F$, where $Fp$ is read as ``the proposition $p$ \textit{will} be true at some future time relative to the present.'' Note that although $F$ introduces temporality explicitely in the syntax, it is not necessarily the case that this leads to a violation of LEM unless one assumes an open-future semantics of a certain kind\footnote{In fact, Prior initially explored systems that allow violations of the law of non-contradiction while retaining the LEM \cite{ploug2012branching}, but he later developed the Peircean and Ockhamist systems \cite{prior2025past, sep-future-contingents}, in which non-contradiction is preserved and the LEM fails specifically for future contingents under branching-time semantics.}

On this note, an important development of Prior’s ideas is the so-called branching-time semantics \cite{belnap2001facing, sep-future-contingents}. In a branching model of time, truth at the present moment is evaluated relative to a set of admissible future branches. At the present time ($t=0$), $Fp$ is true if and only if, for every admissible future branch $b$, there exists a future time $t_b>0$ along that same branch at which $p$ holds. Conversely, $\neg Fp$ is true if and only if, for every admissible future branch $b$ and for every future time $t>0$ along that branch, $p$ fails to hold. When $p$ holds along some branches but not others, neither $Fp$ nor $\neg Fp$ is true at present, and the law of excluded middle $Fp \lor \neg Fp$ fails for such future contingents, even though no third truth value is introduced.

Here we do not commit to any particular logical system; rather, we generically advocate for a two-valued logic in which the law of excluded middle fails for future contingencies, while the principle of non-contradiction is preserved. We take this approach because, in our view, it best captures the notion of creative time in an indeterministic physics (again, this is similar to Aristotle's position \cite{whitaker1996aristotle}, however motivated in naturalistic terms).
Note that such a failure of the law of excluded middle also appears in constructive mathematics---e.g., intuitionism \cite{posy2020mathematical, sep-intuitionism}---making this proposal simultaneously meet our desiderata for a logic model and remain consistent with our perspective on mathematics in Sect.~\ref{math}.
In fact, Kripke models for intuitionistic logic can be understood temporally if we view possible worlds as construction stages that track the mental activity of Brouwer’s constructing subject \cite{vanDalen2001intuitionistic}, or as time steps associated with actualization in our naturalistic view \cite{bentzen2025naturalistic}. Note, however, that temporality is not made explicit in the syntax, as it is in tense logic \cite{vanDalen2001intuitionistic}.

Let us analyze in more detail the failure of the LEM in this view. In our indeterministic framework, physical properties will eventually actualize, yielding \emph{eventual bivalence}: either $p$ or $\neg p$ will be true once the relevant (creative) time has occurred. However, this eventual bivalence does not imply that either $p$ or $\neg p$ is already true at the present time.
In intuitionistic logic, to say that $p$ is true at $t=0$ means that $p$ can be proven at $t=0$, using the information available at that time. in the context of indeterministic physics, ``can be proven'' is understood as ``is in fact already determinate'', as argued in greater detail in~\cite{bentzen2025naturalistic}. Accordingly, for the disjunction appearing in the LEM to be true, the proposition $p \vee \neg p$ must satisfy two conditions: (i) either $p$ or $\neg p$ is true, and (ii) it is \emph{determinate} which of the two alternatives holds.\footnote{In standard (i.e., non-naturalistic) intuitionism, this condition is formulated epistemically: $p \vee \neg p$ is true iff (i) either $p$ or $\neg p$ is true, and (ii) it is known, or has been proven, which alternative holds.} 
Since we assume that propositions about the future are, in general, ontically indeterminate, the LEM does not hold for them at the present time. It is precisely the process of determination -- which restores the LEM for present propositions -- that requires creative time. It may also be useful to consider intuitionistic free logic, since it allows to deal with  objects that may not yet exist, such as worlds or branches that did not yet actualized \cite{posy1982free}.


We emphasize again that our approach differs from ``orthodox'' intuitionism, 
for we reject the Brouwerian notion of the ``creative subject'' in intuitionistic 
logic, advocating instead a naturalistic version of intuitionism in which new 
information is created in the universe by natural processes \cite{bentzen2025naturalistic}.\footnote{There is historical evidence that Brouwer discussed the possibility of 
sequences generated by a physical process, e.g., throwing a die 
\cite{van1995hermann, van1999brouwerian}. Although this is a matter of historical 
debate, it seems that Brouwer rejected this possibility, ``probably because it was 
based on the physical world'' \cite{brouwer1981brouwer}.}

In summary, even in logic we see natural connections between different logic models and determinism versus indeterminism. We have argued that classical logic is most compatible with a deterministic worldview and aligns naturally with geometric time, whereas tense logic and (naturalistic) intuitionistic logic—where the LEM fails—provide a better fit for indeterminism and the notion of creative time.

\section{Relation to the philosophy of time}
\label{phil}

In the vast literature devoted to the philosophy of time (see, e.g., \cite{Whitrow, price2009flow, butterfield2012time, le2015espace, le2019qu} and references therein), one finds a main separation into two camps, following the seminal works of J. M. E. McTaggart \cite{mctaggart1988nature}. The two camps are customarily referred to as A-theory and B-theory of time. While we acknowledge that the distinction in A- and B-theories presents a variety of modifications and it may not even encapsulate the most striking features of time in physics (e.g., directionality, openness of the future, objective becoming, etc.), we cannot avoid drawing some parallel with our distinction between creative and geometric time, respectively.\footnote{Admittedly, there are a large variety of A- and B-theories. Here we believe we follow some widely accepted understandings thereof.}

In fact, starting in reversed order, B-theorists regard positioning in time in dyadic relational terms: given two events $E_1$ and $E_2$, it is either the case that $E_1$ is before $E_2$, or simultaneous to $E_2$, or after $E_2$ \cite{pauri2007time}. Clearly, these relations are somehow $\vir$static", i.e., they are not tensed and hold unchanged since ever and forever.

On the contrary, A-theorists uphold the view that characterizes the positioning of events in time with a monadic attribution: given an event $E$, it is either the case that $E$ is present, or past, or future \cite{pauri2007time}.

Concerning the metaphysical status of time, the B-theory aligns more naturally with the spacetime structure of (deterministic) classical physics and relativity theory, where events are organized on a 4-dimensional manifold. Moreover, B-theorists tend to endorse Eternalism, i.e., the doctrine that all events are temporally on an equal footing and their relation of being in the past, present or future of each other merely reflects their placement on the manifold  \cite{sep-time} (but, as we will see, this is not necessarily the case). Clearly, the concept of geometric time as expressed above fits best with the worldview offered by the B-theory camp.

A first point about Eternalism is that they seem to us to tacitly assume that events exist predeterminate, since all events necessarily \textit{are} -- in the static sense of being part of the block universe. As an example, consider two historically well-known events: the extinction of the dinosaurs and the French Revolution. In an indeterministic world, neither of these events was necessary; therefore, their before-after relation, as conceived in B-theories, would supposedly relate events that might never occur. At a remote past time, when both events -- “dinosaurs’ extinction” and “French Revolution” -- were merely potential, their before–after relation was likewise indeterminate, simply because the events themselves were not bound to happen. After the actualization of the dinosaurs’ extinction, but before the actualization of the French Revolution, one can rule out the possibility that the French Revolution is before (or simultaneous with) the extinction of the dinosaurs. However, it remains generally indeterminate whether the French Revolution will ever actualize (even if this indeterminacy may be characterized by a definite propensity \cite{del2023prop}).\footnote{We are not considering here the further complication that, from an evolutionary perspective, if the French Revolution occurs, this entails that humans have evolved, and thus the event must necessarily follow the extinction of the dinosaurs.} Hence, B-theorists, who take the before–after relation as fixed, seem to us to assume that events are bound to occur.

Moreover, we note that the idea of ontic indeterminacy, which gives rise to an open future, is in tension with B-theories taken at face value. In fact, B-theorists deny the existence of the metaphysical properties of pastness, presentness, and futurity. For instance, in a greatly influential work to support a B-theory view in the philosophy of time, D. H. Mellor states that $\vir$nothing really is past, present or future in itself", and that $\vir$tense is not an aspect of reality"  \cite{mellor2002real}. In particular, we stress that B-theorists reject the idea that the future exists in itself. And by denying the existence of the future altogether, they also cannot accommodate the idea of an open future: There is no such a thing as the $\vir$future" and, \textit{ipso facto}, no $\vir$open future".

Hence, geometric time -- which, as we have argued,  is closely tied to determinism in physics -- finds a more natural alignment with B-theory. In this view, there is no ontological distinction between past, present, and future, and nothing genuinely new ever comes into existence: no new information is added to the world.

A-theorists, on the contrary, accept the thesis of \textit{temporalism}, i.e., that propositions change their truth-value over time, and of \textit{temporal disparity}, i.e., the view that there is a metaphysical distinction between past, present and future \cite{deasy2023limited}. This leads to the view that changes really happen:  $\vir$there is a way reality is (now, presently) which is complete [i.e., maximal given what exists (a pure state as physicist say)], but \textit{was} different in the past and also \textit{will be} different in the future"\cite{sep-time}. A-theory allows for a genuinely open future and fundamental indeterminism, which makes it the framework most naturally aligned with the notion of creative time.
 
The biggest disagreement among different A-theories is about which specific kind of temporal disparity one envisions. There are three main proposals: (i) \textit{Presentism} \cite{markosian2004defense, prior2003papers, wuthrich2012fate}, which maintains that $\vir$no objects exist in time without being present" \cite{sep-time}; (ii) the \textit{Moving spotlight} theory \cite{cameron2015moving}, an eternalist A-theory which takes events to be set on a fix manifold, but the present is singled out as if it was lit by a light moving along the manifold; and (iii) the \textit{Growing-block} theory \cite{forbes2016growing} (sometimes referred as the \textit{Growing-universe}, or \textit{Growing-past}), according to which events come into existence at present but, unlike in presentism, they persist in the past. 

The moving spot-light, as an eternalist theory, does not seem compatible with creative time, showing that there is not a one-to-one correspondence between A-theories and the concept of creative time. 
The view we endorse here is akin in spirit to the Growing-block view, though with important distinctions.
 The growing block states that the present is metaphysically privileged because it is then that events become determinate (or $\vir$real" in the standard parlance of philosophy of time, which we do not endorse). But as new events become present, the past ones remain equally $\vir$real", resulting in a growing block of reality \cite{sep-time}. In particular, we find inspiration in a variant of the Growing-block theory, namely, E. Barnes and R. Cameron's \textit{growing cloud of determinacy theory} \cite{barnes2009open}, which upholds that the openness of the future has nothing to do with its alleged non-existence (a common conclusion in philosophy of time). In the words of the authors, $\vir$the future is as yet unsettled. [T]hink of this ‘unsettledness' with respect to future states of the world as a type of indeterminacy. For all times $t_1$ and $t_2$ such that $t_2$ is later than $t_1$, it is indeterminate at $t_1$ what the state of the world is at $t_2$" \cite{barnes2009open}. As such, this theory aligns well with our physics-inspired idea that the future exists, metaphysically different from the past, as a collection of real potentialities and the actualization of these potentialities is the act of creation that defines creative time.

However, the Growing-block picture, along with the growing cloud of determinacy theory, is tenable only if we assume the existence of an absolute global present -- a statement that holds in classical physics but is rejected by the theory of relativity. Therefore, our version of a growing-block universe incorporates the necessary modifications to remain compatible with relativistic principles. 
The present is defined, for us, as the edge of the events that turn from potential (the open future) to actual (the settled past).\footnote{Although we have discussed the possibility that the remote past could again become indeterminate in Ref. \cite{santo2023open}.} For our view to remain compatible with relativity (see Section~\ref{rela} and Ref.~\cite{del2021relativity}), we must relax the absoluteness of determinate facts: the actualization of a propensity should itself be a relative property, defined with respect to regions of space–time, yet without invoking observers. Hence, a bit that becomes determinate does so only within its future light cone, while it remains indeterminate elsewhere.
 This results in multiple local presents, each defined by the collection of determination events lying on the edges of their past light cone. The set of local presents, however, contrarily to the Growing-block picture, cannot be collected into a global present.

We note that these considerations resonate with an approach to quantum gravity due to R. Sorkin, named \textit{causal sets} \cite{sorkin1991spacetime, dowker2014birth, dowker2021recovering}. This is an approach that takes spacetime as fundamentally made of discrete events related by partial order relations which encapsulate causality. The dynamics is a $\vir$growth process" in which new elements come into existence from their causal ancestors.

Yet, our view allows to maintain that past, present and future are all in a sense real, metaphysically distinct, but relative to a light cone: the past is the (growing) collection of physical properties that got already actualized, the future is the collection of the potential properties,\footnote{Note that some future properties have propensity one, hence are already actual \cite{del2023prop}.} and the present is the transition from potential to actual  (see also \cite{del2023prop}). Similar ideas  were already put forward by H. Reichenbach who wrote: $\vir$The present, which separates the future from the past, is the moment when that which was indeterminate becomes determinate, and 'becoming' means the same as 'becoming determinate'." \cite{reichenbachfond}. In Reichenbach, however, the present is still relative to an observer or to a reference frame; that is, any moment can serve as the division between the determinate past and the indeterminate future, relative to whoever is simultaneous with that event (see \cite{mccall1976objective} and references therein for a discussion).

Our proposal also has several  similarities with S. McCall's $\vir$objective time flow" \cite{mccall1976objective}. Therein, the author considers that the universe has, at any point in time, a dynamical tree structure, with a single determinate past and multiple possible futures. The concept of becoming is brought about by the update of the tree structure, when only one branch becomes determinate. However, McCall's proposal remains at an abstract level, whereas we present a naturalistic (physical) account of the indeterminate-actualization transition. Moreover, McCall does not conceive the possibility of objective (possibly biased) tendencies or propensities towards one possible future or another.

We find it worth positioning our contribution within the metaphysical debate on time. First of all, the novelty of our approach is to provide a naturalistic characterization of the passage of time and of the present, as an objective transition from the past to the future. Moreover, one of the main criticisms against the Growing-block theory (the so-called \textit{skeptic objection}), which rendered this approach quite unpopular among philosophers, is that since both past and present have the same ontological status (i.e., they exist, as opposed to the future), how can we know that we live in the present and not in the past? It seems that our ideas expressed above overcome this skeptical objection exactly due to its naturalistic character, based on indeterministic physics. In fact, although both past and present are  determinate (and therefore different from the future), they are also different from each other. The present is when creative time processes, namely, when the  determination (or actualization) happens and  new information gets created. The present has a dynamical nature, whereas the past remains fixed.

 To conclude, we have shown that geometric time is more naturally aligned with the worldview of B-theories, in which all events exist as well-determinate within a block universe. By contrast, creative time fits best with A-theories, which allow for an open future and thus fundamental indeterminism.

\section{More insights on time(s) from physical theories}
\subsection{Relativity theory}
\label{rela}
As already recalled, geometric time is the conception of time derived from the regularities of some phenomena described by classical physics, like the motion of planets and satellites in the solar system which in turn led to a dominant deterministic world-view in physics. As remarked by  G. E. M. Anscombe, $\vir$the high success of Newton's astronomy was in one way an intellectual disaster: it produced an illusion... for this gave the impression that we had here an ideal of scientific explanation; whereas the truth was, it was mere obligingness on the part of the solar system, by having had so peaceful a history in recorded time, to provide such a model." \cite{anscombe1971causality}. 

We agree with H. Reichenback that $\vir$the properties of time which the theory of relativity has discovered have nothing to do with its treatment as fourth dimension. This procedure was already possible in classical physics, where it was frequently used. However, according to the theory of relativity the four-dimensional manifold is of a new type; it obeys laws different from those of classical theory" \cite{reichenbach2012philosophy}. And indeed it was the advent of the theory of (both special and general) relativity that popularized further the idea of geometric time. Therein,  not only is time a parameter, but it seems to lose any special conceptual role with respect to space (besides the signature of the Minkowski metric, which assigns opposite signs to spatial and temporal components).

\begin{table*}[]
\centering
\resizebox{.65\textwidth}{!}{%
\begin{tabular}{|
>{\columncolor[HTML]{FFFFFF}}c ||
>{\columncolor[HTML]{FFFFFF}}c |
>{\columncolor[HTML]{FFFFFF}}c |
>{\columncolor[HTML]{FFFFFF}}c |
>{\columncolor[HTML]{FFFFFF}}c |}
\hline
Time      & Physics       & Mathematics     & Logic      & Philosophy \\ \hline\hline
Geometric & Determinism   & Classical       & With LEM & B-theory   \\ \hline
Creative  & Indeterminism & Intuitionistic & Without LEM     & A-theory   \\ \hline
\end{tabular}%
}
\caption{Summary of the relation of concepts of creative and geometric time to differnt disciplines.}
\label{table}
\end{table*}

 Geometric time intervals are not relativistically invariant. Indeed, the (geometric) time of a moving observer at speed $v$ (with respect to a given reference frame) undergoes a dilation of a factor $\sqrt{1-\frac{v^2}{c^2}}$, according to the Lorentz transformations. Geometric time can be manipulated by using the scenario of the twin paradox.
This means to prepare two identical systems ($K$ and $L$), send one of them (say $L$) away, and finally bring $L$ back together with $K$. Since $L$'s local clock passes at a slower rate than the one of $K$, the geometric time elapsed for the system $L$ will be shorter than the one elapsed for $K$ (the higher $v$, the larger the dilation). Let us consider again the example of the weather in Geneva in a month's time, and let us assume that the same computation of the outcome is carried out with a computer $K$ and with a moving computer $L$ (assuming that they are using the same algorithm and identical computers). If at the event in which they meet again, $K$'s clock has registered a time $\Delta t'$ (which is the time that it takes for her computer to return an outcome), $L$'s program would still be running due to time dilation. This shows again that geometric time can be manipulated. On the contrary, it is unclear whether there is any way to let creative time $\vir$run slower".

The  question then naturally arises whether creative time, too, transforms relativistically. Looking at concrete experiments from particle physics, one would initially think that it necessarily does.  Indeed, subatomic quantum particles like muons, undergo spontaneous decay which supposedly involves a genuine indeterministic process, and hence creative time. Muons have a mean lifetime of a only 2.2 \textmu s, so on average they should be able to travel a half-survival distance of only about 456 meters without decaying \cite{rhodes2015muon}. The fact that we are able to detect cosmic rays muons (i.e., produced in the upper atmosphere) at sea level, however, means that they exist undecayed for a period (as seen from the reference frame of Earth) that is 25 times longer than their lifetime. Such an effect is explained by the fact that due to the muon's velocity close to the speed of light, the muon lifetime undergoes relativistic time dilation. This could be considered a strong argument to show that also creative time transforms under the Lorentz transformations.

Yet, one can also think that the muon evolves deterministically while propagating through the atmosphere (thereby involving only geometric time) into a state of  quantum superposition of decayed-undecayed and the process that actualizes the outcome $\vir$decayed" (which involves creative time) only happens inside the detector, hence at rest with Earth, 
similarly to an atom in a superposition of an excited and a ground state. If the atom $\vir$decays" (gets de-excited) then a photon is produced and the detection of the latter can demonstrate that the measurement took place. But if a photon is not detected, the atom can still be in superposition until it interacts with a detector. 
 In this way, creative time would not necessarily follow  relativistic transformations. While this is admittedly a wild conjecture, it seems not to contradict the experimental evidence. This calls for further investigations.

\begin{figure}[ht]
\includegraphics[width=.3\textwidth]{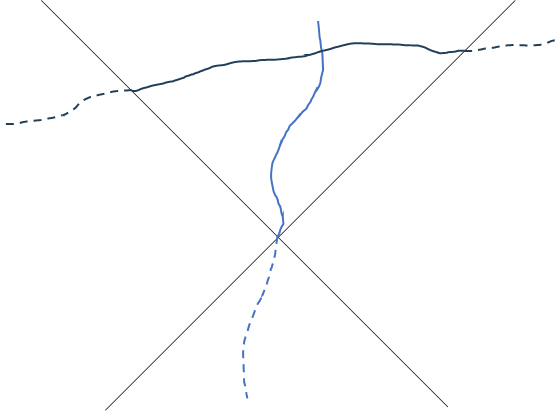}
    \caption{Illustration of determinacy propagation within the light cone. Dotted lines indicate states of indeterminacy, while solid lines represent determinacy.}
    \label{relativity}
\end{figure}

One might think that geometric time is more friendly to relativity than creative time (for a  discussion and proposed counterarguments see also \cite{bondi1952relativity, del2021relativity}). 
But, actually, creative time might also be quite friendly to (special) relativity. Consider a true random number generator that produces (creates) a bit and let us fix the coordinate system such that this bit is produced at a certain spacetime location, say the origin for simplicity. Along a time-like line passing through the origin, there is a time when the bit's value is indeterminate and a time when it is determinate. Next, consider a space-like line that intersects the future light cone (see Fig. \ref{relativity}). Assuming that the bit has a determinate value only inside the future light cone (see \cite{del2021relativity}), one can “move" along that space-like line and notice, similarly to moving along the time-like line, a segment thereof where the bit's value is indeterminate and a segment where it is determinate.
Note that along the time-like line there are only 2 regions --one of indeterminacy and one of determinacy --, while along the space-like line there are 3 regions, i.e., one can “move" from indeterminate to determinate and (without “returning") again to indeterminate.

Finally, the idea that creative time fundamentally separates the past from the future may \textit{prima facie} seem at odds with relativity. However, the events of actualization that characterize the occurrence of creative time are to be considered local therefore there is no (global) simultaneity of the present, in accordance with relativity. In Ref. \cite{del2021relativity}, we have flashed out a proposal to make relativity and fundamental indeterminism compatible. This is to assume that event of actualization from indeterminate to determinate exists only within the forward light-cone, while remaining indeterminate in any other region of spacetime \cite{del2021relativity}.  (In)determinacy is, therefore, itself relative, i.e., determinate facts are not absolute. This idea has gained considerable relevance in recent analyses in the foundations of quantum theory \cite{bong2020strong,brukner2020facts,del2025wigner}. Our proposal, however, develops a theory of relative (determinate) facts with respect to regions of spacetime rather than to observers, as is the case in most of the above references.  To understand this, consider for example two distant, space-like separated, observers, Alice and Bob, each with a local natural  random   number generator. If Alice's random number generator actualizes a previously indeterminate bit to, say, the value $a=0$. For Bob, however, the value of the bit $a$ remains indeterminate until the intersection of the future light cones of Alice and Bob. In this light, the metaphysical property of being determinate is objective, but relative to a certain region of spacetime.

Therefore, different regions have different ticking of creative time. However, one can conjecture that due to quantum nonlocality (of the propensities that characterize the indeterminacy),  different $\vir$local presents" may be synchronized by the actualization at a distance of a nonlocal propensity (see also \cite{del2023prop}).
This admittedly remains a serious challenge and likely calls for a fully relativistic treatment of quantum nonlocality.

\subsection{Quantum theory}

Quantum mechanics is the first physical theory in which indeterminism has been widely -- though not universally --  regarded as fundamental. Indeed, Heisenberg uncertainty principle (which would be better named ``indeterminacy"' principle) and, later on, the violation of Bell's inequalities provided strong indications that properties of quantum systems do not exist predeterminate \cite{pironio2010random}. In this, quantum physics has inspired us to develop the concept of creative time, which happens at the moment when a quantum event, e.g. a measurement, determines a physical property (i.e. a single real outcome) that was previously only potential (as encapsulated by the probability amplitudes in the quantum state) \cite{del2023prop,arntzenius1995indeterminism}. 

This can, however, also be applied in hindsight to classical theory that can consistently be interpreted as a fundamentally indeterministic theory \cite{del2019physics, gisin2021indeterminism}.

Creative time thus becomes associated with how, and under what circumstances the potentialities become actual, i.e. to the notorious measurement problem (also at the classical level, see \cite{del2019physics, del2023prop}). In fact, the concept of creative time is compatible only with these classes of interpretations that conceive a passage from the potential to the determinate \cite{mariani2021indeterminate}. In quantum theory these are the $\vir$spontaneous collapse theories" or interpretations, à la Copenhagen, where the collapse is  induced by (some features of) the measurement. Other interpretations -- such as the many-world and Bohmian mechanics -- cannot accommodate the concept of creative time.

It should be noted that, while indeterminacy in classical and quantum theory leads to some common features, it also presents fundamental differences (see Ref. \cite{del2024features} for a detailed discussion). In quantum mechanics, certain physical quantities are incompatible (represented by noncommuting operators), and their indeterminacies are linked through Heisenberg’s uncertainty relations—for example, position and momentum. 
In contrast, while classical physics can be supplemented with noncontextual hidden variables (such as the real numbers used in the theory), any completion of quantum physics would require contextual and nonlocal additional variables. In philosophical terms, classical indeterminacy is considered shallow, whereas quantum indeterminacy is deep \cite{miller2024classical, del2024features}.

\subsection{Thermodynamics}

A long standing problem that would be impossible not to mention is why do we ubiquitously  observe a direction of time while the microscopic laws of physics are time-reversal invariant. This is commonly explained through statistical consideration, i.e., the fact that the thermodynamic quantity entropy cannot decrease in a closed system (second law of thermodynamics)  
\cite{albert2001time}. This tension between the observed asymmetry in time and the  time symmetry of the underlying microphysics are exemplified by the Loschmidt Paradox and by the Zermelo objection that Boltzmann's H-Theorem is at odds with Poincaré recurrence theorem.
However, this is not necessarily the case even at the classical level, as discussed in detail in the works of B. Drossel and G. Ellis \cite{drossel2015relation, drossel2017ten, drossel2023passage, ellis2020emergence, ellis2024physical}. 

In a similar fashion, our distinction between creative and geometrical time can help clarify this matter. If there are fundamentally indeterminate events that  require creative time, then there is also a fundamental asymmetry between past and future, as encapsulated by the second law. So, while admittedly this still requires to address the measurement problem (classical or quantum), paradoxes like the ones of Loschmidt or Zermelo are solved by merely noticing that  Poincar\'e recurrence theorem simply fails \cite{del2019physics}.
The consequences of indeterminacy -- and the role of creative time -- in statistical mechanics, arguably a  domain where we possess a physical explanation of time’s direction, require further analysis and are the subject of ongoing work.


\section{conclusions}
We join T. Maudlin when he states that $\vir$it is a fundamental, irreducible fact about the spatio-temporal structure of the world that time passes" \cite{maudlin2007metaphysics}, or, in our jargon, that time occurs. However, the formalization of physical theories has to a large extent expelled the real passage of time from the description of natural phenomena (see also \cite{smolin2013time}). We have shown that to understand physics in indeterministic terms -- and we argued that there is strong conceptual and empirical ground to do so -- we should rethink our concept of time and distinguish between a geometric time (that appears in the deterministic equations of motion) and a creative time (that processes concurrently with the events of actualization of potentialities). 

 In physics, indeed, we have tried to relate these two conceptions of time to, respectively, determinism and indeterminism. We then showed a parallel in classical and intuitionistic mathematics. The former takes a Platonistic approach by assuming that mathematical entities (such as real numbers) are given all at once, therefore avoiding the necessity of creative time. On the other hand, we showed that a naturalistic interpretation of intuitionistic mathematics requires a concept of creative time \cite{bentzen2025naturalistic}.We have therefore advocated a set of desiderata for a logical framework that seem to best capture indeterminism and creative time: namely, a logic that incorporates temporality (either syntactically or purely semantically) and that, by adopting an open-future semantics, rejects the law of the excluded middle. Finally, we have related our discussion to some philosophical theories of time, showing in particular that geometric time fits best with the worldview of B-theories, whereas creative time with A-theories. Creative time has also led us to conceive a variation to the Growing-block theory, which however overcomes fundamental criticisms by providing a naturalistic account of the difference between past, present, and future. Notably, we showed that to be consistent with relativity, there are many local presents that cannot be collected into a global present. These parallels in different disciplines are summarized in Table \ref{table}.

We conclude by stating that, in our view, physics is not only about sophisticated theories and fascinating technologies, it should also allow one to tell stories about ``how nature does it". But all stories require time \cite{dolev2018physics}. Hence, we would like to rephrase a famous aphorism by French writer F. Rabelais as: ``science without time is but ruin of intelligibility".\footnote{The original sentence reads: $\vir$Science without consciousness is but ruin of the soul".}

\acknowledgments

We thank Clive Aw, Bruno Bentzen, Jeremy Butterfield, Mauro Dorato, Barbara Drossel, George Ellis, Michael Esfeld, Baptiste Le Bihan, Cristian Mariani,  Tim Maudlin,  Andrea Oldofredi, Carl Posy, Valerio Scarani, Christian W\"uthrich and  Mark van Atten,  for useful discussions and for pointing out relevant literature. This research was supported by the FWF (Austrian Science Fund) through an Erwin Schr\"odinger Fellowship (Project J 4699), and the Swiss National Science Foundation via the NCCR-SwissMap.

\bibliography{biblio}

\end{document}